# Structure transition in PSS-lysozyme complexes: a chain conformation driven process, as directly seen by Small Angle Neutron Scattering

*Jérémie Gummel, Fabrice Cousin*[*] *and François Boué*

Laboratoire Léon Brillouin, CEA Saclay 91191 Gif-sur-Yvette Cedex France

[*] Corresponding author

Fabrice.Cousin@cea.fr

Measurements of chain conformation in proteins/polyelectrolytes complexes (lysozyme and PSSNa) show that the crossover observed between an open structure -a chain network crosslinked by the proteins, and a globular one - dense globules of ~ 10 nm aggregated in a fractal way, results from a conformation modification prior to the transition. Before showing this, we have widened the parameters range for the observation of the transition. We had shown before that the two structures can be formed depending on chain length (for a given [PSS]/[lysozyme] ratio): gel for large chains, globules for short chains. We show here that the crossover between these two regimes can also be reached as a function of the chains concentration or the salinity of the buffer. Since all these crossover parameters act on chains overlapping concentration $c^*$, we reinforce the idea of a transition from the dilute to the semi-dilute regime, but $c^*$ is shifted compared to pure PSS solutions. In order to understand this, we have measured by SANS the conformation of a single chain of PSS in presence of proteins within the complexes. This is achieved by a specific labeling trick where we take advantage of the fact that lysozyme and hydrogenated PSS chains have the same neutron scattering length density. In the gel structure, the PSS chains keep a wormlike structure as in pure solutions, but their persistence length is strongly reduced, from 50 Å without proteins to 20 Å in average with lysozyme. With this value of 20 Å, we calculate new overlapping thresholds (concentration, mass, ionic strength) in agreement with observed ones. In a second stage, after the globular structure is formed, the PSS chains get a third conformation, no longer wormlike, but more collapsed, within the globules.

**I Introduction**

Science of complexes of polyelectrolytes with proteins in aqueous solvents is presently widening and progressing, prompted by its importance in real life (food, biotechnology and biology, personal care, and physicochemical control of various processes). We focus here on electrostatic complexes, resulting from association between two species of opposite electric charges. Understanding the structure of these complexes is quite relevant. Beyond imaginative pictures extrapolated from information at macroscopic scale [1, 2, 3], progress has been done via techniques at scales characteristic of the structure. We have shown recently in a series of papers [4, 5, 6] that neutron scattering, complemented by imaging for larger scale, using FFTEM (freeze fracture) [7], brings several crucial informations. The very "technique- convenient" system we used is mixture of lysozyme, a model globular protein (positive global charge), with Sodium Poly(StyreneSulfonate), PSSNa (negatively charged). This is convenient because the chains can be labelled by deuteration. The key point of these recent studies was that owing to deuterium labelling for neutron radiation, we could measure separately the scattering from protein, and from polymer – and therefore their spatial arrangement.

We actually observed three different structures, corresponding to three regions of the state diagram established from visual and rheological observation. Among the three, two are probably common to many systems, as inferred from macroscopic or light scattering experiments measurements [1, 2] (the third structure, optically clear, shows lysozyme unfolding by a large PSS to lysozyme ratio, a phenomenon encountered at room temperature with PSSNa only). Among the two general structures, the **first structure** shows chains spreading all over the solvent (like in solution), but connected by the proteins which act as individual cross-links. This structure presents a soft elasticity, hence we call it a gel. Such gel-like structure has been observed within BSA/PDMAC coacervates [8] or with hydrophobically modified poly(sodium acrylate) reversibly cross-linked with proteins [9]. The **second structure** is liquid, but at small scale it is much more heterogeneous than the first one. It is made of primary globules with a narrow size distribution (peak value 10 to 30 nanometers depending on the electrostatic parameters), which contain both proteins and chains, at a rather high compactness. These globules aggregate at higher scale in ramified bunches, with fractal dimension 2.1 like the result of a Reaction Limited Aggregation process. Eventually, very large aggregates precipitate at long times. The local organization in globules is common to many systems involving polyelectrolyte and spheres of opposite charges, including complexes made of proteins and polyelectrolytes [1-2, 10-12] (the higher scale organization, fractal in our case, seems more system dependent as coacervates are often observed in close systems) as well as other systems such as complexes made of polyelectrolyte-neutral diblock copolymers and micelles [13, 14], complexes made of polyelectrolytes and micelles [15] or complexes of polyelectrolytes/inorganic charged nanoparticles [16].

  Obviously gel and globule structures are also very different from many macroscopic points of view (rheology, transport, electrophoretic behaviour, permeability, species access, response to dilution, stability). So it is important in practice to control the transition from one to another. The similarity between the structure of the PSS chains within the gel structure and within pure solutions of PSS chains in semi-dilute regime prompted us to imagine in [4] that the a physical origin of the for the this crossover between the two regimes . We imagined the transition is due to the state of interpenetration of the chains after before interaction with the proteins: if chains are interpenetrated (semi-dilute regime), they could easily be crosslinked by the proteins and form a gel, which in turn prevents them to collapse into globules.

We also showed in However our first measurements [4] show that the gel-globule cross-over limit as a function of chain length **does not correspond** to chain overlapping limit, c*, if we calculate it assuming the same chain conformation than in polyelectrolyte solutions, which has been determined, for example, in [17]. This was in accordance with the fact that the PSS chains were shrunk in presence of the proteins.

This leads us to the A possible explanation is then that the **chain conformation is modified under addition of proteins;** this would in turn change c*. It has been for example shown recently by light scattering and viscosity measurements [18] that for PAA, under addition of oppositely charged micelles (similar in shape to globular proteins), c* is shifted towards higher concentrations by a factor 5. Such influence of the addition of oppositely charged pearl-like objects on the overlapping concentration of charged string-like objects has not yet been described, to our knowledge, by theories or simulations. Although there is now a large literature dealing with the adsorption of a polyelectrolyte on an oppositely charged sphere as a function the chain stiffness, chain length, ionic concentration, particle size and surface charge density (see for example the review of Ulrich et al [19]), the simulations cannot yet involve enough objects to resolve overlapping chain concentration problematic.

Simulations [20,21] show that the binding of spherical macroions on polyelectrolyte decreases when the stiffness of the chain increases. Experimentally, this has been confirmed in [22] by turbidity measurements on micelles/polyelectrolytes and proteins/polyelectrolyte systems. We believe thus that the c* of polyelectrolytes would be less affected by complexation for stiff chains. We recall here that PSS has a very small bare persistence length (~ 10 Å [17, 23-24]). Its c* should thus be strongly influenced by complexation. It has to be noted that a strong increase of salt may lead to release of macroions from the chains due to the screening of electrostatic interactions [21]. In this case the c* of the chains should correspond to the one of the pure polyelectrolyte solutions. Nevertheless, simulations involving several chains or several spheres begin to appear. They lead to dense or open aggregates, resembling the globular or gel-like observed structures [25, 26]. In these simulations these differences in structure depend on the strength of interactions and not on the dilution state of the chains.

In order to check that understand whether the concentration cross-over is directly linked to $c^*$, we propose to measure in the present paper it would be highly interesting to measure directly the conformation of chains in presence of proteins inside the complexes. It happens that this is possible, in our system, using neutron scattering. This is due to some fortunate scattering specificity of the components of our system: non deuterated PSS and lysozyme have the same neutronic contrast, so they can be both matched by the same $H_2O/D_2O$ mixture. Meanwhile, deuterated PSS is not matched and still visible in this mixture. This particularity will enable us to determine the form factor of the individual chain.

In the sake of consistency between our former work and the present one, the content of the paper is two-folded:

- one fold reports results using the former method: by contrast matching of either polymer or protein, we probe the gel–globule transition over a larger range of chain concentration and ionic strength. This part being not principal here, some of the data will be presented in Appendix.

- the second fold reports on an original method: the direct measurement of the conformation of the PSS chains in both the gel structure and the globular structure. Results show how that the evolution of chain conformation, first from solution to gel and second to globule, is indeed here the key for the transition.

**II Material and methods**

1 Sample preparation

The synthesis of poly(styrenesulfonate) is done in several steps. We first purchase from Polymers Standards Service poly(styrene) chains of 800 repeating units (Mw = 90000), with very low polydispersity (Mw/Mn = 1.03), in non deuterated and deuterated versions (for contrast matching SANS studies). A post-sulfonation of these chains grafts the sulfonate groups on the aromatic cycles. This is a derivation of Makowski's method [27]. A reactive species, created in-situ by the reaction between sulphuric acid and acetic anhydride, attacks the aromatic cycle, which grafts the sulfonate group in the para position (not the ortho position due to the chain backbone steric hindrance). The poly(styrenic acid) solution is then neutralized by NaOH to obtain a PSSNa solution. This solution is then dialyzed against deionised water. The dialysis is monitored by conductivity and water is renewed as many times as necessary until reaching the conductivity of pure water (18 MOhm). The solution is concentrated in a rotating evaporator and finally freeze-dried, yielding a white powder that can be stored. The sulfonation rate of PSS chains f can be precisely measured by SANS (not shown here). It is well-known that the value of the correlation peak q* of pure solutions of PSSNa depends on f no a reproducible way when all chains are labeled with respect to solvent. For 0.3 Mol/L, one gets q* = 0.1 Å$^{-1}$ [17, 28, 29].

Lysozyme is purchased from Sigma and the same lot has been used for all our studies, without further purification. All samples are made in an acetic acid/sodium acetate buffer solution to reach a pH of 4.7. The buffer concentration is always firstly set to have an ionic strength of $5.10^{-2}$ mol/L. For specific experiments, the ionic strength of the buffer is increased to $1.10^{-1}$ mol/L, $2.10^{-1}$ mol/L or $5.10^{-1}$ mol/L by addition of NaCl.

Two solutions, one of lysozyme and one of PSS, are first prepared separately in the acetic buffer at twice the concentrations wanted. For the measurements of the PSS chain conformation, the PSS solution contains hydrogenated non deuterated and deuterated PSS chains, mixed together in the ratio $PSS_h/PSS_d$ wanted, before mixing with the protein. The two solutions, PSS and protein, are then quickly mixed and slightly shaken to be homogenized. The samples are then left for 2 days at rest; we checked in previous experiments that this is enough to reach a stable state. We define a charge ratio introduced noted $[-]/[+]_{intro}$ which is obtained as a function of the concentrations introduced, taking for the net charge of the lysozyme a value of +11 (at pH 4.7) and for the charge of the PSS one negative charge per monomer. This ratio thus corresponds to the structural charges and not to the effective charges predicted for free chains by Manning's condensation. The charge ratio $[-]/[+]_{intro}$ used here is generally here 3.33. (40 g/L of protein and 0.1 mol/L of PSS repetitions units). $[-]/[+]_{intro}$ is varied only in experiments for which the concentration of PSS chains is changed at constant concentration of protein (40 g/L, Figure A.2).

2 SANS experiments

SANS measurements were done either on PAXY and PACE spectrometers (LLB, Saclay, France) in a q-range lying from $6.10^{-3}$ to $3.10^{-1}$ Å$^{-1}$. All measurements were done under atmospheric pressure and at room temperature. For SANS experiments concerning the determination of the threshold of the transition between the gel and the aggregates of globular complexes regimes, hydrogenated protein and deuterated PSS have been used, like in our previous experiments [4, 5]. Each sample was achieved either in a 57%/43% $H_2O/D_2O$ mixture that matches the neutron scattering length density of lysozyme to get the PSS signal, or in a 100% $D_2O$ solvent that matches the neutron scattering length density of PSS to get the signal of the lysozyme. For the measurements of the conformation of the PSS chains, hydrogenated protein and a mixture of hydrogenated and deuterated PSS have been used. In order to get only the signal of deuterated PSS chains, each sample was made in a 57%/43% $H_2O/D_2O$ mixture that matches both the neutron scattering length density of lysozyme and hydrogenated PSS. The table 1 recalls all the scattering density length of species and solvents used in the different experiments.

[Table 1]

Standard corrections for sample volume, neutron beam transmission, empty cell signal subtraction, detector efficiency, subtraction of incoherent scattering and solvent buffer scattering were applied to get the scattered intensities in "absolute units" ($cm^{-1}$).

**III SANS from protein or polymer: more parameters for the gel to globule transition.**

Before to show the form factor measurements in Section IV, we report in this Section III new data for SANS from protein (polymer matched) or polymer (protein matched) generalising our former study [4, 5]. As before, we consider cases where the charges brought by each component $[-]/[+]_{intro}$ are of the order of 1 (between 0.5 and 3.33).

1 Results

   1.a Effect of chain length.

We first recall here former results on the influence of the chain length [4, 5]. In this work, we had kept all other parameters constant, namely a concentration of 0.1 M for the PSS chains and 40 g/L for lysozyme ($[-]/[+]_{intro}$ = 3.33), and a ionic strength of $5.10^{-2}$ M. These former results [4] are presented in figure 1, together with new results detailed just below. The signals of proteins (matching PSS) and of PSS chains (matching proteins) are plotted separately. Differences between short and long are clear cut:

  - for small chains (N = 50 [5], N = 90 and N = 360 [4], spectra are characteristic of aggregates of globular structures. Lysozyme scattering and PSS scattering have the same features because both species objects are embedded in the globules. We get a strong correlation peak at 0.2 $Å^{-1}$ corresponding to contact distance between two proteins, a second $q^{-4}$ behaviour at intermediate q corresponding to the form factor of the primary compact complexes and a $q^{-2.1}$ behaviour at low q corresponding to the larger scale fractal organisation of the primary complexes (please note that a 2.5 exponent was first reported in [4], but later measurements at low q [5] enabled us to refine the value to 2.1).

  - for longer chains (N = 625 [4]), the behaviour is drastically different. The lysozyme and PSS scatterings are no longer similar. The protein spectra show that proteins are distributed randomly at a small scale and heterogeneously at higher scale with a fractal dimension ($D_f$ = 2.5) akin to spatial distribution of cross-linkers observed in some cases in gels [30]. The PSS chains spectra show that the transient network formed by PSS chains alone in solution remains because the chain-chain correlation peak at intermediate q is still visible. The combination of those two signals enables to describe the structure as a gel where proteins cross-link the PSS network [4]. Macroscopically the sample with N = 625 was a weakly turbid gel whereas the samples with lower chains length were liquid and highly turbid.

In order to check whether the chain length is really a key parameter to shift from one structure to another, we have used a new chain length (N = 800) in the same conditions for the other parameters. Macroscopically the sample is a weakly turbid gel. Both the scattering of lysozyme and PSS chains are plotted in Figure 1 together with the former data. Data for N=800 are similar to the one previously observed for the N = 625 sample. In particular, one recovers the $q^{-2.5}$ decreasing law at low q values in the protein signal and the correlation peak of the PSS network chains in the PSS scattering. This confirms that we have two types of structure dense globules for short chains and crosslinked gel for long chains, as a function of the chain length, with a threshold between N = 360 and N = 625 for the concentrations and ionic strength indicated above.

[Figure 1]

1.b Other parameters.

In order to determine whether this threshold depends on other parameters we have then kept the PSS chains at N = 800 units, and varied the concentration in PSS, from 0.015 M to 0.08 M (we also keep the protein concentration at 40 g/L and an ionic strength at I = 5 $10^{-2}$ M). The complete series of SANS spectra can be found in Figure A.1 in the Appendix 1 section, where it is compared with the 0.1M sample measured formerly. $[-]/[+]_{intro}$ lies between 0.5 and 3.33. The collection of spectra signals gel above a PSS concentration $c_p$ = 0.08 M, and globules below 0.04M. The transition, seen for 0.05 M, occurs on a narrow range.

Finally a last parameter has been tuned, the ionic strength of the buffer, keeping fixed PSS chain length (N = 800) and species concentrations ($[-]/[+]_{intro}$ = 3.33). Four samples have been made starting from 5 $10^{-2}$ M up to 5 $10^{-1}$ M. The SANS spectra (Appendix 2, Figure A.2) show that the two regimes can be reached when tuning the ionic strength, with a critical value clearly between 1 $10^{-1}$ M and 2 $10^{-1}$ M (for this PSS chain length and species concentration).

2 Discussion on threshold values.

What is driving the transition towards globules? Since the three parameters influencing the structure may affect the regime of dilution of the chains, let us revisit our idea that the two structures correspond to the dilute regime of the chains for the globules and to the semi-dilute regime of the chains for the gel. The threshold between those two regimes would then be set by the overlapping critical concentration of the chains c*. As noted in [4], the arrangement of the PSS chains in the gel structure is very close to the one of a pure solution of PSS chains in semi-dilute region. The only observed difference lies in the mesh size of the network formed, which is larger in presence of proteins, suggesting a local shrinking of the chains by the proteins: the q* peak abscissa, corresponding to the network mesh size, is shifted towards low q in presence of proteins.

To calculate the c* of a system of chains with the same wormlike conformation as in a pure PSS solution, we require the radius of gyration $R_g$ of wormlike chains

$$R_g = \sqrt{2Ll_p} \qquad (1)$$

where L is the length of the stretched chain and $l_p$ the persistence length. In case of charged polymers such as PSS, $l_p$ depends both on the intrinsic flexibility of the chain (10 Å for PSS [17, 23-24]), and on electrostatic repulsions between monomers, which depend in turn on ionic strength.

The volume fraction of the solution occupied by the chains is then:

$$\Phi = R_g^3 \times \frac{cN_a}{N} \qquad (2)$$

with c the monomeric concentration of the chains, $N_a$ the Avogadro number and N the number of monomers per chain. The critical concentration c* between the semi-diluted and the diluted regime is reached when the volume fraction is 1 when the chains begin to overlap and thus writes:

$$c^* = \frac{N}{R_g^3 N_a} \qquad (3)$$

The critical concentration finally depends on the number of monomers on the chain (N) and on the persistence length ($l_p$) only.

$$c^* = f\left(\frac{1}{N^{1/2} l_p^{3/2}}\right) \tag{4}$$

Let us now calculate c* for our system and compare to the experimental thresholds. For the case (figure A.1) of a fixed chain length of N = 800, the fixed ionic strength of I = 50 mM determines $l_p$ for PSS chains alone in solution: the value is 50 Å [17] (the counterions coming from PSS and proteins are taken into account in the ionic strength). This yields c* = 0.02 M, which does not match our experimental threshold of 0.08 M. Similarly, if we take the example with variable chain length at fixed PSS concentration, 0.1 M, and ionic strength, I = 50 mM (data of figure 1) we find, using the same formula and $l_p$ = 50 Å, a chain length threshold N* = 30, which would put all our samples in the semi-diluted regime.

In summary, for the two cases, the c* of PSS chains calculated with the persistence length corresponding to pure PSS solutions does not match the gel-globule threshold. But in the mixture the chains interact with the proteins that bear opposite charges. A first effect was suggested by the shrinking of the PSS network (larger q* observed in [4]). In view of Equation (4), the essential parameter is the persistence length, which could be modified after interaction of PSS with the proteins. By measuring the conformation of the PSS chains inside the complexes, it will be possible to determine if it is still wormlike, and to obtain a new value of the persistence length in presence of proteins in order to calculate a new critical concentration c*.

**IV The Conformation of the PSS chains**

In order to get the conformation of the PSS chains inside the complexes in a SANS experiment, one needs to separate in the scattering signal their form factor from the strong interchain structure factor arising from the electrostatic repulsions between the chains. The first solution to be imagined is to work in very low concentrations to weaken as much as possible the structure factor. But changing the dilution would act on the conformation of the chains and the measurement would be senseless. A tricky method is thus required to get rid of interchain scattering, and keep only intrachain contribution. The method we choose consists in making a set of several different samples with same total chain concentration but different fractions of deuterated and non deuterated chains in a solvent matching the scattering length density of hydrogenated chains. When the amount of deuterated chains tends to zero, the system tends to behave as a set of single PSS chain from a scattering point of view, thanks to the hiding of the hydrogenated chains. But at the same time the interactions between the chains in the system do not change because the total concentration of chains remains identical. The principle is recalled on the sketch of figure 2.a that shows a system of pure PSS chains in semi-dilute regime for the four ratios of deuterated chains we chose (25%, 50%, 75% and 100%) in a $H_2O$ solvent (upper sketch) and in a 57%/43% $H_2O/D_2O$ solvent (lower sketch) that matches the hydrogenated chains. In general, this method would not apply for chains within complexes because the protein scattering would remain. Here comes the great advantage of our system: the neutron density length of lysozyme is equal to the one of the hydrogenated PSS! So both species can be matched simultaneously and the measurement of conformation can be done also in presence of proteins (figure 2.b) (see the sketch of figure 2.b).

[Figure 2]

For pure solutions of PSS chains, since we have now two kinds of PSS chains, hydrogenated ones and deuterated ones, mixed in the solution, the total scattering takes into account the correlations between chains of the same kind and chains of different kind:

$$I(q) \, (cm^{-1}) = (1/V) \cdot d\Sigma/d\omega = I(\vec{q}) = \frac{1}{V} \left\langle \sum_{i,j} k_i k_j \exp(i\vec{q}(\vec{r}_i - \vec{r}_j)) \right\rangle \quad (5)$$

where $k_i$ (cm or Å) = $b_i - b_s (V_{mol\_i}/V_{mol\_s})$ is the "contrast length" between one repeating unit of scattering length $b_i$ and molar volume $V_{mol\_i}$, and a solvent molecule ($b_s$, $V_{mol\_s}$).

Let assume first that all chains are labeled in the same way with respect to solvent; i.e. for all i, we have $k_i = k_A$. The concentration is $c_p$, in mole/L (or mole/Å$^3$), so the total volume fraction of chains is $\Phi_T = N_a.c_p.V_{mol\_i}$, where $N_a$ is the Avogadro number.

$$I(q) \, (cm^{-1}, \text{ or } \text{Å}^{-1}) = (1/V) \cdot d\Sigma/d\omega = k_A^2 \, S_{TAA}(q) \quad (6)$$

Using Å and Å$^{-1}$ as the units for $k_H$ and $I(q)$, we obtain $S_{TAA}(q)$ in Å$^{-3}$. Quite generally,

$$S_{TAA}(q) = S_{1AA}(q) + S_{2AA}(q), \tag{7a}$$

where

$$S_{1AA}(q) \, (\text{Å}^{-3}) = \frac{1}{V} \left\langle \sum_{\substack{\alpha \text{ with} \\ \beta = \alpha}} \sum_{i,j} \exp(i\vec{q}(\vec{r}_i^{\,\alpha} - \vec{r}_j^{\,\beta})) \right\rangle \tag{7b}$$

corresponds to the correlations between monomers i,j of the same chain $\alpha = \beta$ (intrachain scattering), and

$$S_{2AA}(q) \, (\text{Å}^{-3}) = \frac{1}{V} \left\langle \sum_{\substack{\alpha, \\ \beta \neq \alpha}} \sum_{i,j} \exp(i\vec{q}(\vec{r}_i^{\,\alpha} - \vec{r}_j^{\,\beta})) \right\rangle \tag{7c}$$

corresponds to the correlations between monomers i,j of two different chains $\alpha$ and $\beta \neq \alpha$ (interchain scattering).

Assume now that chains are labeled in two different ways. In practice, we use a mixture of d-PSS chains ($k_i = k_D$), and h-PSS chains ($k_i = k_H$). The scattered intensity becomes:

$$I(q) \, (\text{cm}^{-1}) = (1/V) \cdot d\Sigma/d\Omega = k_H^2 \, S_{HH}(q) + 2 \, k_H \, k_D \, S_{HD}(q) + k_D^2 \, S_{DD}(q) \tag{8}$$

with $S_{HH}(q)$ the scattering of the hydrogenated chains, $S_{DD}(q)$ the scattering of the deuterated chains and $S_{HD}(q)$ the cross-term. In a solvent that matches the scattering length density of the hydrogenated chains, $k_H = 0$, hence

$$I(q) \, (\text{cm}^{-1}) = (1/V) \cdot d\Sigma/d\omega = k_D^2 \, S_{DD}(q) \tag{9}$$

The total volume fraction of chains in the solution is the sum of the volume fractions of the two types of chains, $\Phi_T = \Phi_H + \Phi_D$. Since H and D chains are perfectly identical except for the value of b (in particular $V_{molH} = V_{molD} = V_{mol}$), we can write:

$$S_{DD}(q) = \phi_D \, S_1(q) + \phi_D^2 \, S_2(q) \tag{10}$$

$S_1(q)$ and $S_2(q)$ concerning all H and D chains being defined as above by (7b) and (7c). Eventually

$$I(q) \, (\text{cm}^{-1}) / \phi_D = k_D^2 \, (S_1(q) + \phi_D \, S_2(q)) \tag{11}$$

If we add now lysozyme in the system, numerous new correlation terms are to be taken into account! But in the 57%/43% $H_2O/D_2O$ solvent that matches here also the protein neutron density length (as well as the hydrogenated chains), all terms with $k_H$ in the front factor vanish, a huge simplification which makes equation (11) still valid.

Experimentally, we will measure, for a given system, the scattering for four $[PSS_d]/[PSS_h]$ ratios. For each q value, the four intensities will be linear as a function of $\Phi$, and the extrapolation to $\Phi=0$ will provide the value of $S_{1D}$. The dimensionless form factor of the PSS chains, P(q) ($\lim_{q\to0} P(q) = 1$), will finally be obtained from $S_1(q)$:

$$\lim_{\phi_D \to 0} I(q)/(\phi_D \cdot k_D^2) = S_1(q) \, (\text{Å}^{-3}) = (1/V_{mol}) \cdot N_w \cdot P(q) \qquad (12)$$

at each q values probed in the experiment.

In the following section we describe three measurements of PSS form factors, one for a solution of pure PSS chains and two for solutions of PSS chains complexed with proteins in the gel-like structure and in the globular structure.

1 Results

1.1 Pure PSS solutions.

We have firstly determined the conformation of the PSS chains alone in solution to check whether we find with this method of measurement the value of the persistence length given in the literature (which was actually obtained for the first time with the same method [24]). We use a solution of PSS at 0.1 M which corresponds to the concentration we will use to determine the conformation inside the complexes, and a buffer ionic strength of $5.10^{-2}$ mol/L. The rate of deuterated chains $\Phi_d$ ($[PSS_d]/[PSS_h]$) is set to 25%, 50%, 75% and 100%. The figure 3 shows all the scattering spectra renormalized by $\Phi_d$ to enable a direct comparison between the signals.

[Figure 3]

For high q values (q > 0.05 Å$^{-1}$) the four signals are equivalent but at low q values they become sensitive to the structure factor. In this low q-region, one gets a correlation peak due to the strong electrostatic repulsions between the chains. At q tending towards zero, the structure factor decreases to values much lower than one. The contribution of the correlation peak on the scattering progressively vanishes when the amount of deuterated chains decreases, and the signal progressively tends to the signal of one chain. The extrapolation at $\Phi_d = 0$ at each q value as explained before is represented in red in Figure 3 [31]. At low q, it is thus higher than all the measured ones. At high q, it is similar to the experimental scattering signals values (q > 0.05 Å$^{-1}$) because in this q region, the structure factor is very close to 1. The extrapolated intrachain scattering is presented in a Kratky plot (Iq² = f(q), linear axes, q range from q = $10^{-3}$ to q = $6.10^{-2}$) in Figure 4, which enables a good comparison of the experimental signal with the calculated scattering of a wormlike chain [32].

**[Figure 4]**

In this model, the wormlike chain has a constant curvature characterized by a persistence length $l_p$, characterised as follows. If $\Psi$ is the angle between the tangents of two points of the chains separated by a distance l, then:

$$<\cos(\Psi)> \approx \exp(-l/l_p) \qquad (13)$$

Sharp and Bloomfield [33] have given the following expression for the form factor P(q) of wormlike chains that takes into account the finite size of chains of length L, at low q :

$$P(q) = \frac{2(\exp(-x) + x - 1)}{x^2} + \left[\frac{4}{15} + \frac{7}{15x} - \left(\frac{11}{15} + \frac{7}{15x}\right)\exp(-x)\right]\frac{2l_p}{L} \qquad \text{for} \qquad qL_p < 4 \qquad (14)$$

with $x = \frac{Lq^2l_p}{3}$. This expression is valid until $qL_p < 4$. It enables us to fit all the part of the signal at the lower q values. When $qL_p > 4$ then the signal follows an asymptotic law of Des Cloizeaux [34]:

$$P(q) = \frac{\pi}{qL} + \frac{2}{3q^2Ll_p} \qquad (15)$$

The molecular weight distribution in our measurements (in other words the chain length L), are known. The extrapolated spectrum is here in absolute intensities (cm$^{-1}$), and we calculate the volume fraction $\phi_D$ of the chains from their molar volume $V_{mol}$, their concentration (0.1M) and the neutronic contrast ($\Delta\rho^2 = k_D^2/V_{mol}^2 = 1.49 \; 10^{21}$ cm$^{-4}$). The low q value enables to check that one recovers the molecular weight Nw of the PSS chains [35]. The only one parameter that can be tuned to fit the experimental signal is the persistence length $l_p$. It is the sum of the intrinsic persistence length which is due to the chemical rigidity chains (~10 Å) and the electrostatic persistence length due to the repulsion between the charged monomers [36]. On Figure 3 is represented the best fit using equation (12) and (13), yielding $l_p$ of 50 Å, equal to values formerly published for the same ionic strength [17, 24].

1.2 PSS inside the gel structure

The pure solution case being successful, we have used the same protocol with a PSS – lysozyme mixture in the gel regime. The concentration of PSS is 0.1 M, the concentration of lysozyme is 40 g/L, which leads to an introduced charge ratio [-]/[+]$_{intro}$ of 3.33. The buffer ionic strength is $5.10^{-2}$ mol/L. A macroscopic gel is obtained for every sample. The amount of deuterated and hydrogenated chains does thus not affect the structure. On the figure 5 are represented the four PSS signals for $\Phi_d$ = 25%, 50%, 75% and 100% as well as the extrapolated form factor [31].

[figure 5]

The signals obtained at the different $\phi_D$ are very close to the ones of pure PSS solutions. The main differences are the position of the polyelectrolyte peak which is shifted towards the lower q, and a slight upturn at low q. The slight upturn is probably due to small heterogeneities in the network of PSS chains. The shift of the peak was observed formerly [4] and attributed to a shrinking of the PSS network mesh. Like in the case of PSS chains alone, it vanishes when $\Phi_d$ tends to 0 and disappears in the extrapolated signal at $\Phi_d$ = 0, shown in green on the figure 5. The latter is also plotted in a Kratky representation in figure 4, for comparison with the signal of pure PSS solution at the same concentration. In the gel structure, PSS chains still display a wormlike behaviour, but their persistence length has changed, as obvious from:

- the different q value for the upturn of the plateau appearing in the Kratky plot: the upturn position is at q = 0.025 Å$^{-1}$ for the PSS inside the complexes when it is at q = 0.015 Å$^{-1}$ for the PSS alone.

- as well as from the different ordinate of this plateau: $1.2.10^{-19}$ cm$^{-3}$ for the PSS inside the complexes whereas it is $5.10^{-20}$ cm$^{-3}$ for the PSS alone.

These strong changes indicate that the persistence length has decreased. The same conclusion can be drawn from a fit to a wormlike chain, yielding $l_p$ = 20 Å instead of 50 Å for pure PSS solutions. The chains are shrunk.

1.3 PSS inside the globular structure

Finally, we have determined the intrachain scattering of PSS inside the globular complexes. For a direct comparison with the two other cases, we used the same chain and lysozyme concentrations ([PSS] = 0.1 M, [lysozyme] = 40 g/L). In order to force the structure to be globular, we increased the ionic strength of the buffer to 0.5 M, on the basis of results of figure A.1. The four PSS signals for $\Phi_d$ = 25%, 50%, 75% and 100% as well as the extrapolated intrachain scattering, are represented in figure 6 [31].

**[Figure 6]**

This time, all signals are very different from the ones from pure PSS solutions. First, the scattering of the 100% d sample, which indeed is the scattering from polymer only, as already seen in Figure 1: whereas at high q values it still decays with a typical $q^{-1}$ decreasing law, at slightly lower q, a peak at 0.2 Å$^{-1}$ due to the contact of the proteins is present here. This correlation is clearly seen in globular complexes for the protein-protein signal, but it is here also seen in the PSS-PSS scattering. At smaller q, the scattering decays like $q^{-4}$; this is the surface scattering of the large dense globules (radius larger than 300 Å).

If we look now at the different samples with lower deuteration rates, these two features (the peak at 0.2 Å$^{-1}$ and the $q^{-4}$ intermediate behaviour) are progressively vanishing because they are both linked to the structure factor. Hence, unlike the two precedent cases, the intensity of the signal at low q values is higher in presence of the structure factor: interactions are here no longer repulsive but attractive, since PSS chains are aggregated within the globules. When the deuteration rate tends to zero, the low q intensity tends towards the scattering of one chain, and thus decreases. Note that all the different signals cross at the same abscissa, q =0.02 Å$^{-1}$. This means that the structure factor is null for this wave vector, so we get a direct measurement of the absolute intensity of the form factor for any $\phi_D$. At high q values, the $\Phi_d$ = 0 extrapolated intrachain scattering is close to the signals of pure PSS and PSS inside the gel structure: it scatters like $q^{-1}$, with a crossover when going towards low q values.

This intrachain scattering factor is compared in the figure 4 with the form factor of the pure PSS determined previously in a Kratky plot. In this representation it is possible to see immediately that the conformation adopted by the PSS chains inside the dense complexes is completely different from the two previous ones. Chains are no longer wormlike. Instead, the $q^2I(q)$ displays a slight maximum around 0.025 Å$^{-1}$ followed at larger q by a plateau.

This $q^2I(q)$ plateau suggests a $q^{-2}$ scattering behaviour showing chains close to a Gaussian behaviour, following a random walk inside the globules. The simplest model to be imagined to fit the plateau is a Debye function corresponding to the form factor of a Gaussian chain with the gyration radius of the PSS chain (57 Å for N = 800). Such a calculated curve (not shown here) would be very different from the experimental one because the plateau would start around 0.05 Å$^{-1}$ instead of 0.02 Å$^{-1}$ here, and its ordinate would be ~ 4 10$^{-19}$ cm$^{-3}$ instead of 1.5 10$^{-19}$ cm$^{-3}$ here. This is in accordance with the fact that the inner volume of the globules is not fully accessible to the chains because of the presence of the proteins. The chains have thus an effective gyration radius higher than the one of a pure Gaussian chain. Moreover a part of the chains would be located in a PSS shell around the globules.

A calculation of a form factor corresponding to a linear combination of a fraction of chains in Gaussian state with the effective radius, and a fraction of PSS wormlike chains is presented in appendix 3. The effective radius of gyration is estimated by calculating the number of proteins complexed per chain. This enables us to fit correctly the value of the plateau in absolute scale (see figure 4) at large q but not the onset of the plateau (0.02 Å$^{-1}$) neither the smooth maximum. One may wonder whether this maximum is linked to the size of the globules: this is not the case since it would give for the globules a radius of 250 Å, much lower than their mean radius which is at least 300 Å for a salinity buffer of 0.5 M, as extracted from the PSS-PSS or protein-protein signal (these scatterings - figure A.1- still vary like $q^{-4}$ at q = 0.0035 Å$^{-1}$, i.e. R >1/0.0035 Å> 300 Å).

An alternative could be found in a third type of combination, between wormlike chains and spheres: the maximum in $q^2S_1(q)$ chain signals at large scale a globular shape, while at short scales (at large q) it is not very different from a rod-like behaviour (in practice it smoothly crosses the wormlike chain signal). Such combination was found for partially sulfonated PSS chains, when some of the units are hydrophobic (like polystyrene). In this case, the pearl necklace structure predicted by Dobrynin and Rubinstein [37] was observed [29]. In the case studied now, our maximum is at ten times lower q, corresponding to spheres larger than 100 Å, but, again, as said just above, they are somewhat too small compared to the globule size, as extracted from the PSS-PSS or protein-protein signal. Please also note that the complexation between both components still occur at large salinities, contrary to the release of proteins at high I suggested by simulations [19,21].

**V Discussion**

**Values of c\* and gel-globule threshold.** The measurement of the conformation of the PSS chains within the gelled samples shows that the interactions between the proteins and the polyelectrolyte leads to the shrinking of the PSS chains, as we suggested in [4], because the persistence length of the chains is strongly reduced, from 50 Å in pure PSS chains solution, down to 20 Å in presence of proteins. Please note that this value of 20 Å is likely to be an average value. Some chains parts, not linked, could keep their genuine 50 Å persistence length, which suggests that for other parts, $l_p$ would be even shorter (the minimum is the intrinsic value 10 Å).

Let us make use of this new value of the persistence length to test if the transition between the two regimes (gel structure or dense globular structure) can be the transition from the dilute regime of the chains to the semi-dilute one. We recalculate the overlapping concentration of the chains, with $l_p = 20$ Å. From Equation (4), the critical chain polymerisation degree for the dilute to semi-dilute crossover at unit concentration 0.1 M (and I = 50 mM) is $N^* = 500$. This new value perfectly matches the gel to globule experimental threshold, $360 < N^* < 625$ (see figure 1).

We can also recalculate c\* for a given chain length at a fixed ionic strength. We obtain c\* = 0.08 M for N = 800 monomers at I = 50 mM, which also perfectly matches the gel-globule threshold found by SANS in appendix 2. This value of c\* is also very close to 0.1 M, chain concentration of the samples for which the buffer ionic strength was tuned in appendix 1, where the transition appears for I between 100 and 200 mM. From Equation 4, at these chain concentration and length, we obtain for a transition between 100 mM and 200 mM, $l_p^* = 17$ Å, a value lower by 3 Å when passing from 50 mM to 150 mM.

In summary, the calculation of the c* with the new value of the persistence length, whatever the parameter tuned (PSS chain length, PSS concentration, ionic strength), strongly confirms that the dilution regime of the PSS chains is the key factor for the transition between the 'gel' structure and the globular structure.

**Chain conformations**. A closer look at the conformation of the PSS chains enables us to understand how the two structures are formed. In the gelled samples, after interaction with the proteins, the key point is that some parts of chains remain wormlike. The form factor shape is compatible with a combination of two conformations. This suggests that some parts of the chains are not complexed with the lysozyme and remain free in solution. For the samples studied in part IV (average $l_p$ of 20 Å), a simple calculation gives 75% of the chains collapsed on proteins (with $l_p = 10$ Å) and 25% free in solutions (with $l_p = 50$ Å). For [PSS] = 0.1 M, this gives 0.025 M of free chains, still higher than c* for $l_p = 50$ Å when the chains are long enough (0.02M for N = 800 and 0.022M for N = 600). The free chains can thus still form the same transient network as in pure PSS solutions. The transient network of the free PSS sequences is thus crosslinked by the chain sequences bound to lysozyme.

For the globular samples, the complexation can be considered as a two-steps process. An initial complexation of the chains with the proteins leaves 25% of the chains free for [PSS] = 0.1 M and [lysozyme] = 40g/L. When the concentration of these free chains is too dilute to form a network (ones gets for example a c* of 0.031M for N = 360 to compare to a concentration of free chains of 0.025 M), the system collapses to form the globular structure due to the electrostatic interactions between species of opposite charge and the gain of entropy associated to the release of counterions [6]. In such a structure, the core of the globules is very dense and electrically neutral, suggesting a local interaction of all PSS units with a positive charge of lysozyme [5]. The PSS chains are thus strongly confined by proteins and cannot thus adopt a wormlike behaviour anymore. In $q^2I$ plot, the slight maximum maybe due to a spherically collapsed global shape, while the plateau observed at large q suggests that the inner conformation is not far from a random walk inside an effective volume fixed by the number of proteins interacting with the chain (see appendix 3).

**VI Conclusion**

We have shown that the transition between the two main structures of PSS/lysozyme complexes, aggregated globules and gel, corresponds to chain overlapping. Prior to this, we have to account for a first stage where chain spatial expanse is reduced by interaction with the proteins. The consecutive concentration threshold c* can then be tuned by usual parameters: chain length, tuning of the persistence length by the salinity of the buffer. The control of the transition is especially important for tuning properties such as rheology, response to dilution, or access to enzymatic sites for applications in pharmacology or biosensors.

This conclusion is made possible by the measurement of the form factor of chains within the complexes. We describe the chain conformation as wormlike, semi-flexible, we extract a persistence length and we show that it decreases in presence of proteins after electrostatic attractions between species. To our best knowledge, this is the first form factor measurement for a polymer in a mixed system of two interacting species in aqueous solutions. It is based on a Small Angle Neutron Scattering method, which is powerful but sophisticated: it requires two different labelled versions of the polymer (here hydrogenated and deuterated PSS) and that the second component of the mixture has the same neutron scattering length density as one of the labelled version of the polymer (here lysozyme and hydrogenated PSS). Obviously, this is not always possible in complexes, and therefore seems demanding. However, many components encountered in soft matter have similar neutron scattering length density in their hydrogenated version and most of common polymers are available in deuterated versions; therefore we would like to point out that this method may apply to other systems.

**Appendix 1: SANS determination of the gel/globule threshold when PSS chains concentration is varied**

In this series of experiment, 6 samples have been prepared with the same amount of protein (40 g/L) at a fixed ionic strength of $I = 5.10^{-2}$ M in a 100% $D_2O$ solvent (matching of the PSS scattering). The concentration of deuterated PSS chains has been varied from 0.015 M to 0.08 M to determine the threshold between the gel regime and the globular regime as a function of the concentration of PSS chains.

[Figure A.1]

The spectra are compared on figure A.2 to the one of the sample at PSS = 0.1 M (see figure 1). For the highest concentrations of PSS chains (0.08 M and 0.1M) the structure formed is the gel (the scattering spectra shows the typical $q^{-2.5}$ decay behavior at low q and the absence of correlation peak at high q). Then for lower concentrations, the scattering decay at low q progressively shifts from a $q^{-2.5}$ behavior to a $q^{-4}$ one for the lowest concentrations in PSS chains while the characteristic correlation peak at $q = 0.2$ Å$^{-1}$ of the globular regime begins to appear. The threshold between the gel and the globular structure can therefore be defined to be at a concentration of PSS chains of 0.08 M.

**Appendix 2: SANS determination of the gel/globule threshold when ionic strength is varied**

In this series of experiment, 4 samples have been prepared with the same amount of protein (40 g/L) and of deuterated PSS chains (0.1M) in a 100% $D_2O$ solvent (matching of the PSS scattering). The ionic strength has been varied from $I = 5.10^{-2}$ to $I = 5.10^{-1}$ M to determine the threshold between the gel regime and the globular regime as a function of the ionic strength.

[Figure A.2]

The spectra, presented on figure A.1, show the two different behaviors depicted in part III.1 for the two lowest and the two highest ionic strengths. For the two highest ionic strengths, we recover the typical features of the globules (the correlation peak at 0.2 Å$^{-1}$ and the q$^{-4}$ scattering decay at intermediate q) and the typical features of the gel for the two lowest ionic strength (the q$^{-2.5}$ scattering decay at low q) The threshold between the gel and the globular structure can therefore be defined in the following range of ionic strength: $1.10^{-1} < I < 2.10^{-1}$ M.

**Appendix 3: Attempt of adjustment of the PSS form factor signal in the globular structure**

In order to adjust the signal of the PSS chains inside the dense globules we have assumed that the chains have a Gaussian-like behavior with an effective gyration radius inside the complexes fixed by volume occupied by the number of proteins interacting with one chain. This radius is the main parameter of the model. First of all we estimate the number of proteins $N$ interacting with a chain with the number of monomer $N_{mono}$. The inner charge ratio within the globules $[-]/[+]_{inner}$ can be determined by the ratio of the PSS signal to the protein signal intensities $I_{PSS}/I_{lyso}$ at low q values (the full method is explained in [5]). This value is estimated to be around 2.5 for the chain with 800 monomers at 0.5M (the lysozyme spectra is given in figure A.1 and the PSS spectra in figure 5). As each PSS monomer bears a negative chargen, we get:

$$N = \frac{N_{mono}}{Z_{lyso}[-]/[+]_{inner}} \qquad (A.1)$$

with $Z_{lyso}$ the net charge of lysozyme. For the chain considered here, the number of protein $N$ is 30 proteins. The associated volume occupied is then dependant on the protein compactness:

$$V_{occup} = \frac{N.V_{lyso}}{\Phi_{inner\_lyso}} \qquad (A.2)$$

with $V_{lyso}$ the volume of lysozyme and $\Phi_{inner\_lyso}$ the volume fraction of the proteins within the globules. This latter can be evaluated from the lysozyme scattering modelization (the full method is explained in [5]) and is 0.15. The effective gyration radius of confinement of a chain can then be directly calculated:

$$R_g = \sqrt[3]{\frac{3V_{occup}}{4\pi}} \qquad (A.3)$$

The value obtained is 85 Å. We consider that the chain has a Gaussian-like behaviour within this effective gyration radius. The form factor of the chain is thus a Debye function:

$$P(q) = \frac{2(e^{-X} - 1 + X)}{X^2} \quad \text{with } X = q^2 R_g^2 \tag{A4}$$

In the globular regime, we showed in [5] that the globules are surrounded by a shell layer coexisting with free chains when $[-]/[+]_{intro} = 3.33$. In order to take into account the signal of these free chains, we have considered that they a standard wormlike chain behavior with the persistence length of pure solutions of PSS chains. For this high ionic strength (0.5M), $l_p$ is 30 Å [13]. The final form factor result is a linear combination of signals of Gaussian chains and wormlike chains taking into account the amount of free chains. This amount of free chains has been calculated from the inner charge ratio $[-]/[+]_{inner}$ and is of 23%.

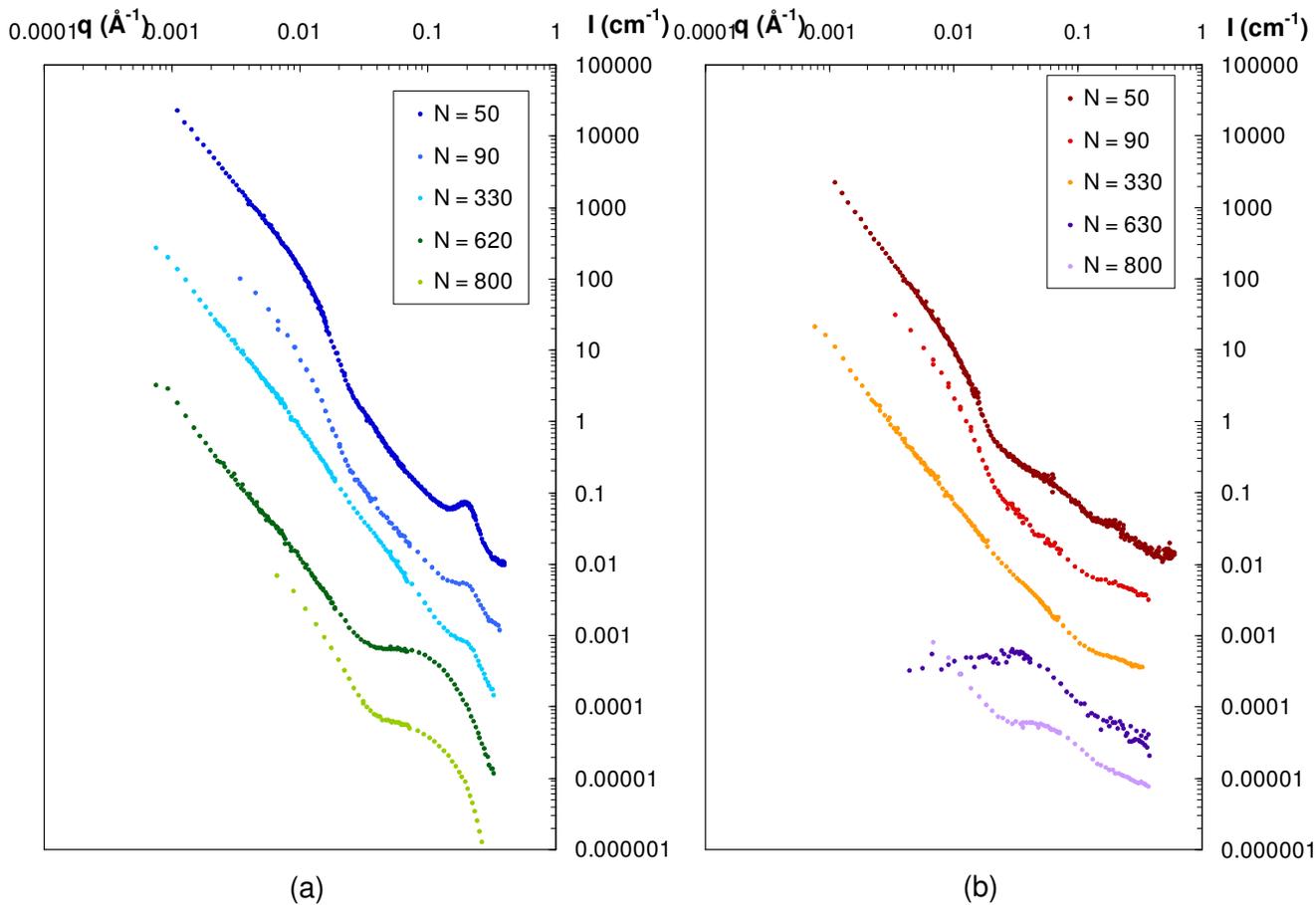

Figure 1 : SANS spectra as a function of the PSS chain size for $[-]/[+]_{intro} = 3.33$ and $I = 5\ 10^{-2}$ M. (a): protein scattering, all the curves are shifted from one to another by a decade for clarity. (b): PSS scattering, all the curves are shifted from one to another by a decade for clarity. The spectra for N=90, N=360 and N=625 have already been published in [4] and the spectra for N=50 have already been published in [5]. The errors bars are smaller than the symbols.

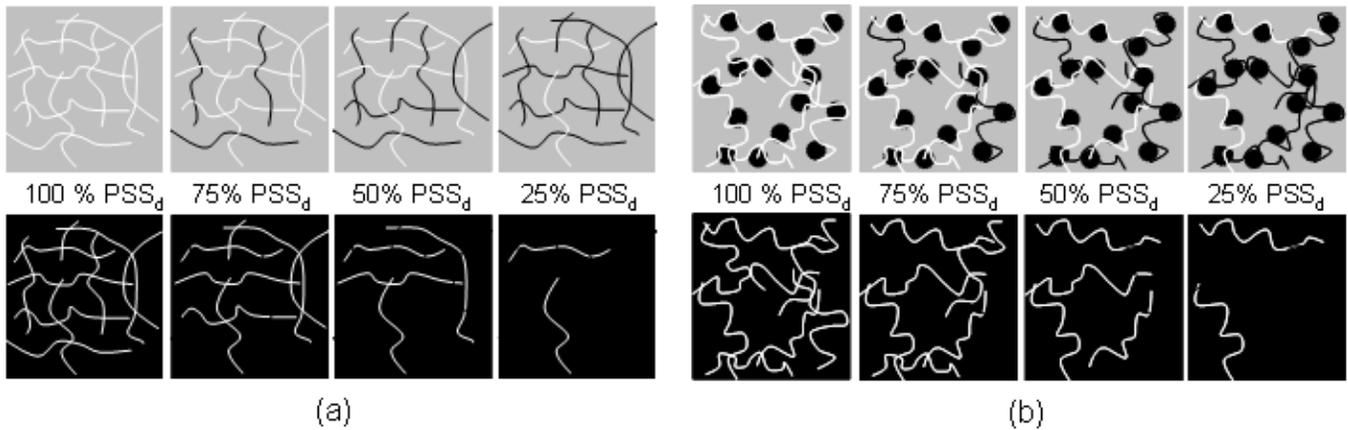

Figure 2 : Principle of the measurement of the form factor of a PSS chain by contrast matching in a SANS experiment. (a) Measurement in pure solutions of PSS chains. (b) Measurement in complexes made with lysozyme. The total amount of PSS chains is constant for all samples. For each drawing, lysozyme and $PSS_h$ chains are both in black and $PSS_d$ chains in white, either in 100%$H_2O$ solvent (grey, upper skecth) or in a 57%$H_2O$/43%$D_2O$ solvent (black, down sketch).

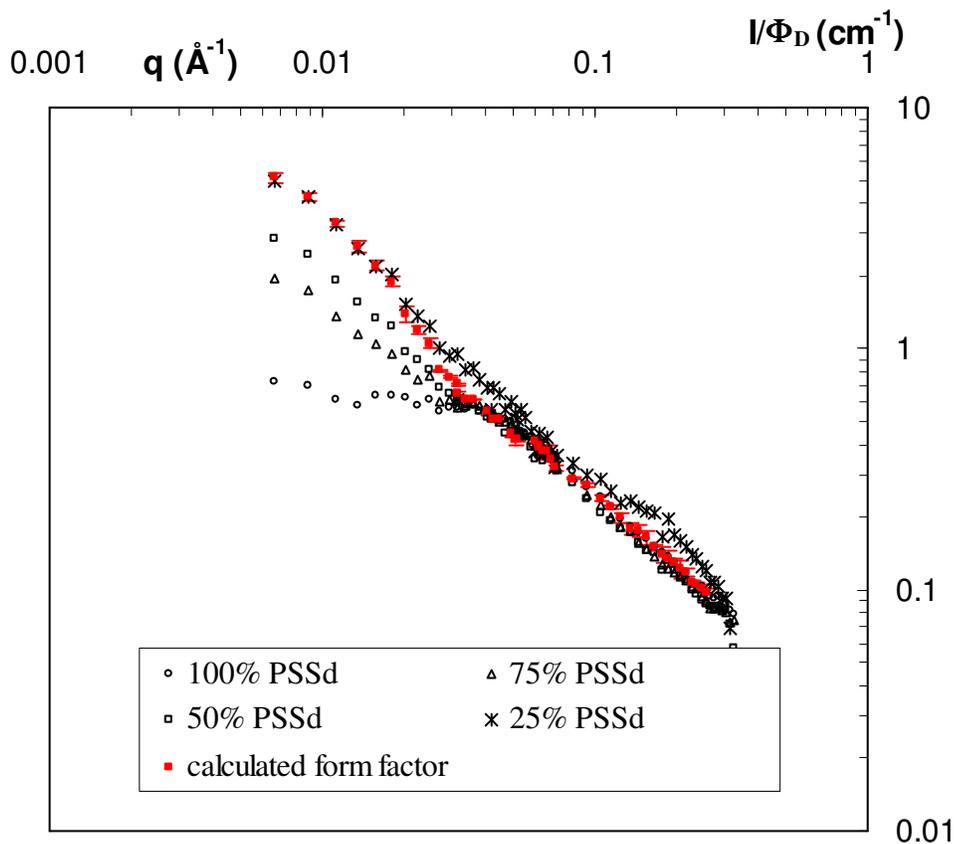

Figure 3. Conformation of a PSS chain in pure solutions of PSS chains ([PSS] = 0.1M, I = 5 $10^{-2}$ M). SANS spectra for the four ratios of deuterated chains in a 57%/43% $H_2O/D_2O$ solvent (open symbols) and extrapolation of the PSS form factor (red filled squares). The errors bars for the experimental data are smaller than the symbols.

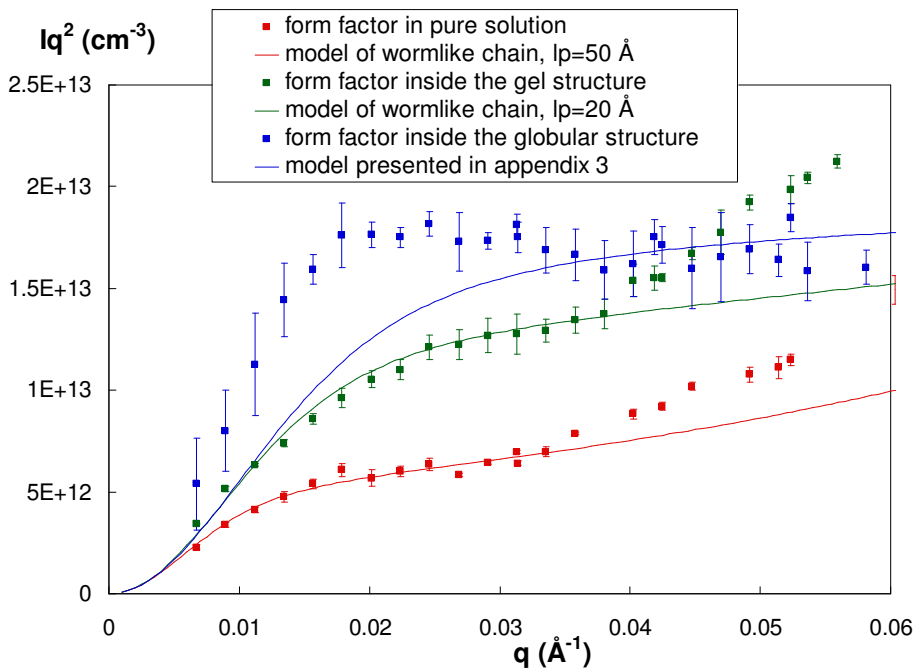

Figure 4 : Kratky plot of the three PSS form factors : Pure solution of PSS (red squares); PSS chains inside the gel structure at I = 5 $10^{-2}$ M (green squares); PSS chains inside the globular structure at I = 5 $10^{-1}$ M (blue squares). The continuous lines correspond to the fits. Red line : form factor of a wormlike chain ($l_p$ = 50 Å ). Green line : form factor of a wormlike chain ($l_p$ = 20 Å ). Blue line : model of a linear combination of a factor of Gaussian chains with an effective gyration radius of 85 Å and of a wormlike chain ($l_p$ = 30 Å ). See appendix 3 for details of the third model.

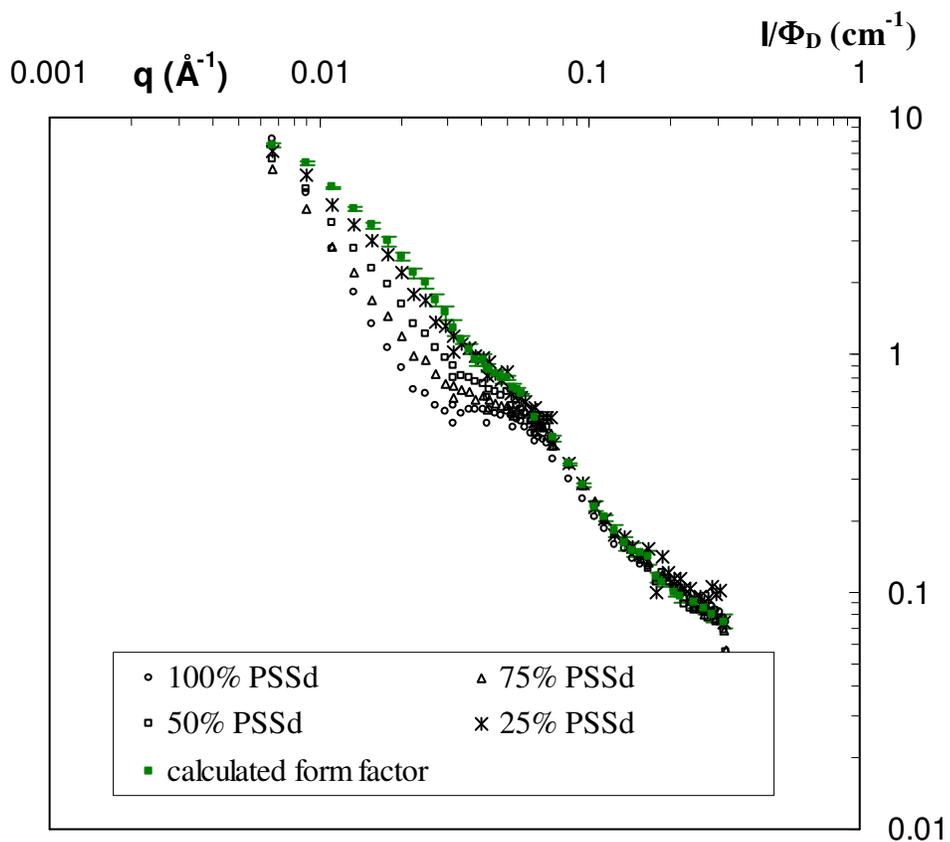

Figure 5 : Conformation of a PSS chain in complexes with a gel-like structure. ([Lyso] = 40g/L, [PSS] = 0.1 M, I = 5 $10^{-2}$ M). SANS spectra for the four ratios of deuterated chains in a 57%/43% $H_2O/D_2O$ solvent (open symbols) and extrapolation of the PSS form factor for the PSS inside the complexes (green filled squares). The errors bars for the experimental data are smaller than the symbols.

Figure 6 : Conformation of a PSS chain in complexes with a globular structure. ([Lyso] = 40g/L, [PSS] = 0.1 M, I = 5 10$^{-1}$ M). SANS spectra for the four ratios of deuterated chains in a 57%/43% H$_2$O/D$_2$O solvent (open symbols) and extrapolation of the PSS form factor for the PSS inside the complexes (blue filled squares). The errors bars for the experimental data are smaller than the symbols.

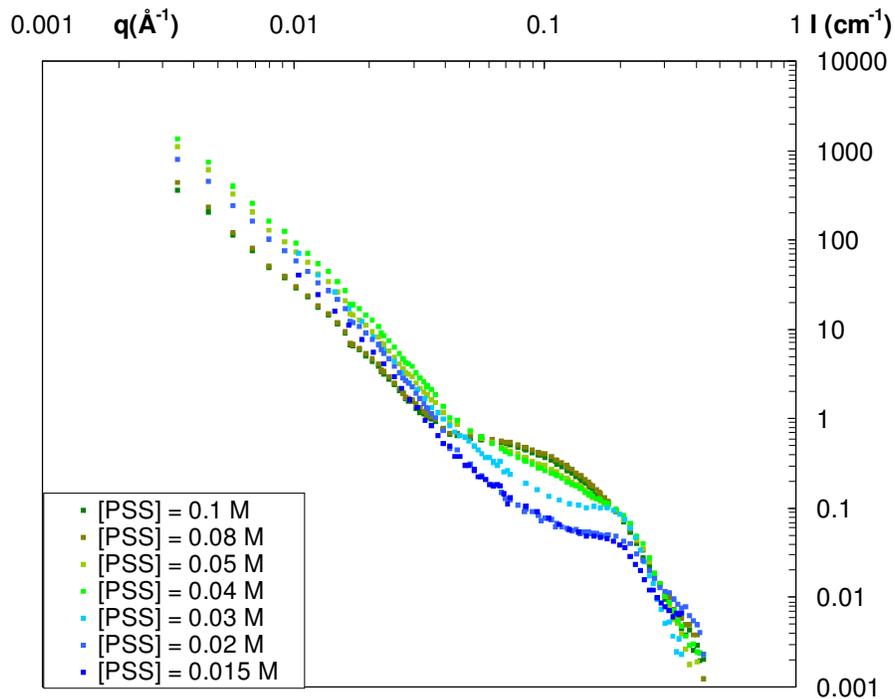

Figure A.1 SANS spectra as a function of the concentration of PSS chains for a concentration in protein of 40 g/L in a 100% $D_2O$ buffer (protein signal) at a fixed ionic strength of I = $5.10^{-2}$ M (proteins scattering). For the two lowest concentrations of PSS, the signal coming from free proteins has been subtracted (see [5] for full details). The errors bars are smaller than the symbols.

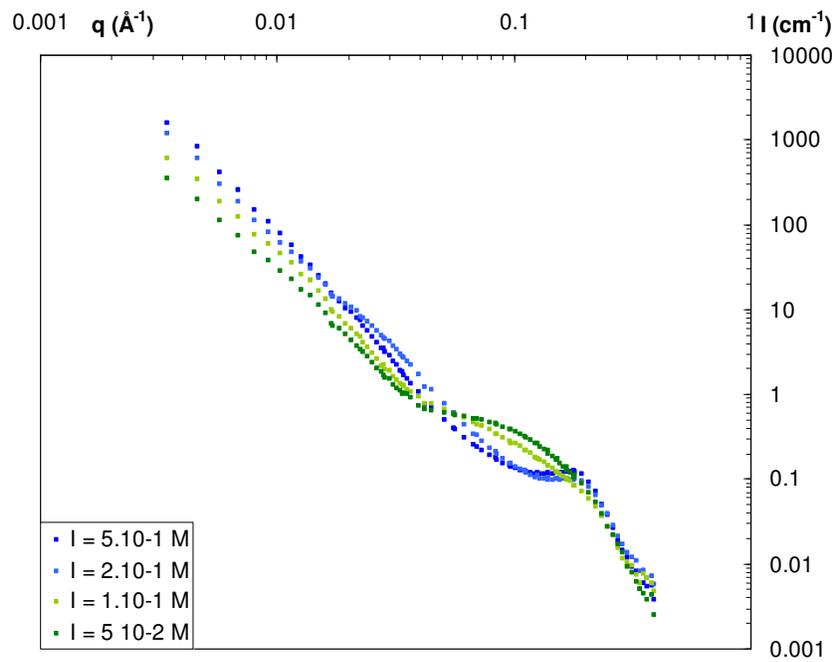

Figure A.2 SANS spectra as a function of the ionic strength at a fixed concentration in PSS chains of 0.1 M and in protein of 40 g/L in a 100% $D_2O$ buffer (protein scattering). The errors bars are smaller than the symbols.

|  | Lysozyme[*] | h-PSS chains | d-PSS chains | 57%/43% $H_2O/D_2O$ | 100%$D_2O$ |
| --- | --- | --- | --- | --- | --- |
| $\rho$ (cm$^{-2}$) | 2.52 10$^{10}$ | 2.52 10$^{10}$ | 6.26 10$^{10}$ | 2.52 10$^{10}$ | 6.39 10$^{10}$ |

[*]This value corresponds to the value in a 57%/43% $H_2O/D_2O$ mixture and takes into account the exchange of the labile hydrogen at the surface of the protein with the solvent.

Table 1 : Scattering length density of the different species and solvent used in the experiment.